\documentclass[useAMS,usenatbib]{mn2e}
\usepackage{natbib}
\usepackage{color,graphicx} % graphicx needed for \includegraphics
\usepackage{amsmath}        % need ams stuff    %
\usepackage{amsfonts}       % to make equations %
\usepackage{amssymb}        % look nice         %
\usepackage{framed}         % for putting framed boxes around text
\usepackage{upgreek}
\usepackage{xspace}
\usepackage{array}
\newcommand{\sarahfrb}{FRB~010125\xspace}
\newcommand{\kb}{FRB~010621\xspace}
\newcommand{\lb}{FRB~010724\xspace}
\newcommand{\tone}{FRB~110220\xspace}
\newcommand{\ttwo}{FRB~110626\xspace}
\newcommand{\tthree}{FRB~110703\xspace}
\newcommand{\tfour}{FRB~120127\xspace}
\newcommand{\spitlerfrb}{FRB~121102\xspace}
\newcommand{\emilyfrb}{FRB~140514\xspace}

\newcommand{\dmunits}{$\mathrm{cm}^{-3}\,\mathrm{pc}$\xspace}
\newcommand{\s}{\textsc{seek}\xspace}
\renewcommand{\d}{\textsc{destroy}\xspace}

\newcommand{\dd}{\textsc{dedisperse$\_$all}\xspace}
\renewcommand{\H}{\textsc{heimdall}\xspace}

\author[Keane \& Petroff]{E.F.~Keane$^{1,2}$ \& E. Petroff$^{1,2,3}$
  \\ $^{1}$ Centre for Astrophysics and Supercomputing, Swinburne
  University of Technology, Mail H29, PO Box 218, VIC 3122,
  Australia. \\ $^{2}$ ARC Centre of Excellence for All-sky
  Astrophysics (CAASTRO). \\ $^{3}$ CSIRO Astronomy \& Space Science,
  Australia Telescope National Facility, P.O. Box 76, Epping, NSW
  1710, Australia \\ } \date{\today} \title[FRBs]{Fast radio bursts:
  search sensitivities and completeness}

\begin{document}

\maketitle

\begin{abstract}

%  Fast radio bursts (FRBs) are millisecond-duration radio signals
%  thought to originate from cosmological distances. Many authors are
%  endeavouring to explain their progenitors, with others outlining
%  their potential uses as cosmological probes. Here we describe some
%  sub-optimal performance in existing FRB search software, which can
%  reduce the volume probed by over $20\%$, and result in missed
%  discoveries and incorrect flux densities and sky
%  rates. Recalculating some FRB flux densities, we find that \sarahfrb
%  was approximately $50\%$ brighter than previously
%  reported. Furthermore we consider incompleteness factors important
%  to the population statistics. Finally we make data for the archival
%  FRBs easily available, along with software to analyse these.
%
  In this paper we identify some sub-optimal performance in algorithms
  that search for Fast Radio Bursts (FRBs), which can reduce the
  cosmological volume probed by over $20\%$, and result in missed
  discoveries and incorrect flux density and sky rate
  determinations. Re-calculating parameters for all of the FRBs
  discovered with the Parkes telescope (i.e. all of the reported FRBs
  bar one), we find some inconsistencies with previously determined
  values, e.g. \sarahfrb was approximately twice as bright as
  previously reported. 
%The inferred burst energy estimates are highly
%  uncertain and span 2 orders of magnitude meaning we do not yet have
%  sufficient precision to determine whether or not FRBs are standard
%  candles. 
  We describe some incompleteness factors not previously considered
  which are important in determining accurate population statistics,
  e.g. accounting for fluence incompleteness the Thornton et
  al. all-sky rate can be re-phrased as $\sim2500$ FRBs per sky per
  day above a 1.4-GHz fluence of
  $\sim2\;\mathrm{Jy}\,\mathrm{ms}$. Finally we make data for the FRBs
  easily available, along with software to analyse these.

\end{abstract}


\section{Introduction}
Fast radio bursts (FRBs) are millisecond-duration jansky-flux density
signals which have been discovered by single dish radio telescopes
operating at frequencies of
$\sim1.4$~GHz~\citep{lbm+07,kkl+11,tsb+13,sch+14,bb14,pbb+14}. All but
one of the 9 reported events have been detected with the 64-m Parkes
Telescope; the other with the 300-m dish at Arecibo. The bursts
exhibit a frequency-dependent time delay which obeys a quadratic form
so strictly that the signals could only have traversed low density
regions, such as the interstellar and intergalactic media, en route to
Earth~\citep{den14}. The magnitude of this delay --- parametrised by
the dispersion measure (DM), which is the integrated electron density
along the line of sight --- is so large that cosmological distances
are inferred for the sources of the
FRBs~\citep{ioka03,inoue04}. Because of this the potential
astrophysical/cosmological uses of FRBs are numerous, and include
weighing the `missing baryons'~\citep{mcq14}, measuring the
intergalactic magnetic field and determining the dark energy equation
of state~\citep{glz14,zlw+14}.

In this paper we present, in \S~\ref{sec:SP_searches}, an assessment
of the search algorithms used to discover FRBs, highlighting some key
concerns where sensitivity to FRBs is often unnecessarily reduced, and
how this can be avoided. In \S~\ref{sec:archival_FRBs} we re-calculate
some basic parameters for all of the Parkes FRBs in a self-consistent
manner. These serve as input to \S~\ref{sec:incompleteness} where we
describe some incompleteness issues in our sampling of the FRB
population. In \S~\ref{sec:last} we then make our conclusions.

\section{Single Pulse Searches}\label{sec:SP_searches}
Thusfar FRBs have been detected in beam-formed radio observations as
follows: (i) \textit{Acquisition:} Radio telescopes 
%\footnote{Thusfar all FRBs have been detected in
%  beam-formed observations. They are potentially detectable also in
%  images if they are highly scattered, e.g. at radio frequencies of
%  $\lesssim 100$~MHz, see \citealt{hkf13}} 
record incoherent filterbank data: these are time-frequency-flux
density data cubes. Thusfar the data wherein FRBs have been discovered
have been centred at $\sim 1.4$~GHz with bandwidths ranging from
$288-400$~MHz, frequency resolutions ranging from $0.336-3$~MHz and
time resolutions ranging from $0.064-1$~ms; (ii) \textit{Cleaning:}
The data are cleaned of radio frequency interference signals in
various ways; (iii) \textit{Dedispersion:} The data are dedispersed at
a number of trial dispersion measure (DM) values to remove
frequency-dependent delays imparted by the interstellar and
intergalactic media; (iv) \textit{Search:} Each dedispersed time
series is subjected to a single pulse (SP) search, which is a matched
filter search to a number of trial boxcar widths. Usually events down
to a level which is well within the noise floor are recorded; (v)
\textit{Refinement:} Upon detection optimised DM and width values of
the pulse are derived.

In steps (ii)--(iv) there is the potential for a loss in
sensitivity. All of these are avoidable, but accuracy is sometimes
sacrificed for processing speed. The DM parameter is always covered in
a `scalloped' fashion, where the next DM trial is chosen so as to
limit the sensitivity loss of a narrow pulse falling between DM
trials. Typically the choice is to lose no more than $\sim10\%$ of the
sensitivity for bursts narrower than the sampling time. However FRBs
are typically much wider than the sampling time, and the observed
width is dominated by dispersion smearing so that the loss in
sensitivity to FRBs is typically much less than this. As DM
corresponds to the volume probed in a line-of-sight dependent way, the
actual volume probed can be quite uncertain, especially for
lines-of-sight closer to the Galactic plane.

The searching step can be subject to the `root 2 problem', which
manifests itself when performing a `decimation search'. This is a
procedure in which a time series is searched for events of 1 sample in
width. It is then down-sampled by a factor of 2, averaging adjacent
samples. This process is repeated a number of times in order to search
for a range of pulse widths~\citep{cm03}. However, this search method
is not optimal. For example, let's consider a time series with samples
i, i=1,2,3,4,..., and a top-hat pulse which is 4 bins wide, occupying
bins 3,4,5 and 6. This pulse is `out of phase' with respect to the
down-sampling procedure and it is clear that the derived S/N will be
too low by a factor of $\sqrt{2}$. The optimal way to search a time
series is to run a sliding boxcar along the time series.

\begin{figure}
  \begin{center}
    % l b r t
    \includegraphics[scale=0.35,trim = 15mm 20mm 15mm 15mm, clip, angle=0]{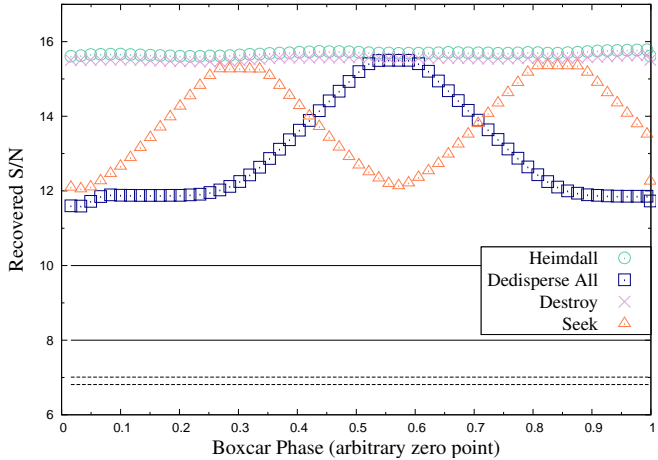}
    \caption{\small{Here we show the detected S/N of a simulated FRB
        (intrinsic pulse width $2$~ms, DM of $1000$~\dmunits observed
        at $\sim1.4$~GHz) for four commonly used SP search codes, as a
        function of `boxcar phase'. The injected S/N in this case is
        $16$. Due to dispersion smearing, and searching for
        power-of-$2$ boxcars the maximum theoretical detection S/N is
        $15.4$. Here we plot the mean recovered S/N values for $100$
        random realisations. We can see that \H (circles) and \d
        (crosses) recover the correct S/N regardless of the position
        of the phase. \dd (squares) and \s (triangles) give S/N values
        which are strongly dependent on phase, with a maximum loss
        factor of $\sim\sqrt{2}$. The rms deviation of the recovered
        S/N values (not shown) are in all cases $\sim 1$, as
        expected. Also shown are various thresholds: a $10$-$\sigma$
        threshold which is often used for real time FRB
        searches~\citep{pbb+14}, an $8$-$\sigma$ threshold more
        commonly used for offline processing, two `false-alarm'
        thresholds for typical pulsar search parameters: the lower
        (higher) dashed line for observations with $2^{21}$ ($2^{23}$)
        samples, $1000$ DM trials and $10$ width
        trials.}}\label{fig:comparison}
  \end{center}
\end{figure}

Figure~\ref{fig:comparison} illustrates this problem by showing the
results of searching for a synthesised FRB as it is moved along a time
series in single time sample steps up to one pulse width in total. The
results of several commonly used SP search codes are shown. In
particular these are
\H\footnote{\texttt{http://sourceforge.net/projects/heimdall-astro/}},
\dd\footnote{\label{footnote_label}see
  e.g. \texttt{https://github.com/SixByNine/psrsoft}},
%\s\footnotemark[\ref{footnote_label}] and
% commented out the above replacing it with below fail soln.
% due to arxiv fussiness
\s$^2$ and
\d\footnote{\texttt{https://github.com/evanocathain/destroy\_gutted}}. Here
a `typical' FRB with an intrinsic pulse width of $2$~ms (16 bins for
the simulated time sampling value of 125~$\upmu$s), a DM of
1000~\dmunits, and an injected S/N of $16$, is used. The data are
centred at a frequency of $1374$~MHz with spectral resolution of
$3$~MHz meaning the dispersion smearing will make the observed pulse
width $7.4$~ms (59 bins). Thus, for a power-of-2 boxcar search the
optimal S/N we expect to find is $16\max(\sqrt{32/59},\sqrt{59/64}) =
16\sqrt{59/64} = 15.4$.

We can see that \dd reaches the maximum theoretical S/N when the pulse
is `in phase' and reaches a minimum S/N when the pulse is `out of
phase'. \s has the same problem, although to a lesser extent, and is
relatively rotated in phase and with a response curve which repeats at
twice the frequency. These differences are because \s performs an
extra 2-bin smoothing step to the data prior to each down-sampling
step. \H and \d give the correct result for all `phases'.
%% Probably reader doesn't care about such tiny details
%The response of \dd is well approximated as $0.76\;(1 +
%0.32\;G(\varphi | \mu=0.57,\sigma=0.13))$ with that of \s being
%$0.78\;(1 + 0.29\; ( G(\varphi | \mu=0.11,\sigma=0.11) + G(\varphi |
%\mu=0.64,\sigma=0.11) )) $, where $G$ is a Gaussian, $\varphi$ is the
%boxcar phase, and with an error of $1$ at each phase.
%are slightly lower, a result of intentionally using an iterative
%approach for determining the time series mean and standard
%deviation. \H uses the absolute mean deviation to calculate these
%quantities, which is not effected by outliers (i.e. the pulse). When
%using this method \d gives identical results to \H.
We have verified that the curves in Figure~\ref{fig:comparison} scale
directly with the injected S/N. One can quickly make a very simplified
estimate of the effects of these results on a survey, e.g. for an
$N\propto S_{\mathrm{min}}^{-3/2}$ law \s probes $86\%$ and \dd only
$78\%$ of the volume probed by \H and \d. The true volume probed will
be slightly lower than this however as here the exact DM of the pulse
has been used, so that the `scallop' loss factor (which would effect
all four codes equally) has been removed. Crucially \textit{this can
  mean that FRBs detectable in our data are never detected}. Some of
these errors can also result in incorrect flux density and volumetric
rate estimates for those FRBs which are detected.

\section{The Parkes FRBs}\label{sec:archival_FRBs}
The 8 Parkes FRBs were discovered using different search software ---
\citet{lbm+07} used \s, \citet{kkl+11} used \d, both \citet{tsb+13}
and \citet{bb14} used \dd, and \citet{pbb+14} used \H. Because of
this, and the issues highlighted in \S~\ref{sec:SP_searches},
previously reported parameters like the observed flux densities,
widths and fluence (and hence derived parameters like the energy
released in the FRB event) may be incorrect. We have therefore
re-determined the signal-to-noise ratios and observed widths of all of
the Parkes FRBs in a self-consistent way, and hence determined the
observed flux densities and fluences. The results of these
calculations are tabulated in Table~\ref{tab:FRB_vals_measured}.

\begin{table*}
  \begin{center}
    \caption{\small{This table summarises some of the measured
        parameters for the Parkes FRBs (where the date IDs have been
        corrected where appropriate), with the reported values for the
        Arecibo FRB also listed for reference. Those FRBs discovered
        using the analogue filterbanks (AFBs, $96\times3$~MHz
        bandwidth) at Parkes are marked with a $^{\dagger}$ and those
        discovered with the Berkeley-Parkes-Swinburne Recorder (BPSR,
        $1024\times0.390625$~MHz bandwidth) are marked with a
        $^{\clubsuit}$. All Parkes events were discovered using the
        13-beam receiver, where the central and most sensitive beam is
        surrounded by two ever less sensitive hexagonal rings of six
        beams~\citep{swb+96}. The beam number wherein each burst was
        detected is given. The published DM values are followed by the
        S/N as calculated by \d. The observed width is that which
        gives the maximum S/N value. In the case of the beam 6
        detection of \lb the 1-bit digitisers saturated so that the
        S/N listed is a lower limit. The observed peak flux density is
        calculated using the listed S/N and width values, a system
        temperature of $28$~K (Parkes Observing Guide, Sep 2014
        Edition), the usable bandwidth after RFI-affected channels are
        removed (conservatively this is $261$~MHz for the AFBs and
        $338.28125$~MHz for BPSR), the relevant digitisation loss
        factor (AFB: $0.798$, BPSR: $0.936$), and the relevant
        beam-dependent gain factor~\citep{mlc+01}. The observed
        fluence is simply the product of the width and the peak flux
        density values. The true peak flux density and fluence values
        are higher than the observed values by $1/G(\theta)$, the
        inverse of the angular response of the beam: the mean of this
        boost factor is $\sim 2$ (J-P. Macquart, priv. comm.). Note
        that the time sampling used varied: \sarahfrb, \kb and \lb
        were observed with time resolutions of $0.125$, $0.250$ and
        $1.000$~ms respectively, whereas all of the BPSR detections
        had time resolution of $0.064$~ms. References correspond to
        [1] \citet{bb14} [2] \citet{kskl12} [3] \citet{lbm+07} [4]
        \citet{tsb+13} [5] \citet{pbb+14} [6] \citet{sch+14}
    }}\label{tab:FRB_vals_measured} \setlength{\extrarowheight}{3 pt}
    \begin{tabular}{ccccccccc}

      \hline\hline Event & Parkes & DM & S/N & $W_{\mathrm{obs}}$ & $S_{\mathrm{peak,obs}}$ & $F_{\mathrm{obs}}$ & Ref.\\
       & Beam ID & ($\rm{cm^{-3}\,pc}$) &  & (ms) & (Jy) & (Jy\,ms) &  \\
      \hline
      Parkes FRBs & & & & & & & & \\
      \sarahfrb$^{\dagger}$  &  5 &  790(3)    & 25(1)  & $10.3^{+2.9}_{-2.5}$ & $0.55^{+0.11}_{-0.08}$ & $5.6^{+3.0}_{-2.0}$ & [1]\\
      \kb$^{\dagger}$        & 10 &  746(1)    & 18(1)  & $8.3^{+4.0}_{-2.3}$  & $0.52^{+0.13}_{-0.11}$ & $4.3^{+3.6}_{-1.9}$ & [2]\\
      \lb$^{\dagger}$        &  6 &  375(1)    & $>100$ & $\sim20$             & $>1.58$                & $>31.5$              & [3]\\
                             &  7 &  375(1)    & 16(1)  & $9^{+12}_{-2}$       & $0.38^{+0.07}_{-0.15}$ & $3.4^{+6.1}_{-1.8}$ & \\   
                             & 12 &  375(1)    & 6(1)   & $33^{+12}_{-28}$     & $0.09^{+0.17}_{-0.03}$ & $2.9^{+8.9}_{-2.6}$ & \\
                             & 13 &  375(1)    & 27(1)  & $15^{+4}_{-3}$       & $0.58^{+0.10}_{-0.08}$ & $8.7^{+4.1}_{-2.7}$ & \\
      \tone$^{\clubsuit}$    &  3 &  944.38(5) & 54(1)  & $6.6^{+1.3}_{-1.0}$  & $1.11^{+1.12}_{-0.10}$ & $7.3^{+2.4}_{-1.7}$ & [4]\\
      \ttwo$^{\clubsuit}$    & 12 &  723.0(3)  & 12(1)  & $1.4^{+1.2}_{-0.4}$  & $0.63^{+0.20}_{-0.13}$ & $0.9^{+1.3}_{-0.4}$ & [4] \\
      \tthree$^{\clubsuit}$  &  5 & 1103.6(7)  & 17(1)  & $3.9^{+2.2}_{-1.9}$  & $0.45^{+0.21}_{-0.10}$ & $1.8^{+2.3}_{-1.1}$ & [4] \\
      \tfour$^{\clubsuit}$   &  4 &  553.3(3)  & 13(1)  & $1.2^{+0.6}_{-0.3}$  & $0.62^{+0.13}_{-0.10}$ & $0.8^{+0.6}_{-0.3}$ & [4] \\
      \emilyfrb$^{\clubsuit}$ & 1 & 562.7(6)   & 16(1)  & $2.8^{+3.5}_{-0.7}$  & $0.47^{+0.11}_{-0.08}$ & $1.3^{+2.3}_{-0.5}$ & [5] \\ \\
      Arecibo FRB & & & & & & & \\
      \spitlerfrb            & n/a & 557(2)    & 14(1)  & $3.0(5)$             & $0.4^{+0.4}_{-0.1}$    & $1.2^{+4.0}_{-1.0}$ & [6] \\

    \end{tabular}
  \end{center}
\end{table*}

We report higher values of S/N and width for \lb and the four Thornton
bursts. For the latter the average reduction in S/N is consistent with
the average of the \dd response curve in
Figure~\ref{fig:comparison}. It is certainly conceivable that other
FRBs, especially weaker events, were missed in the \citet{tsb+13}
search, which motivates a complete re-processing of the high Galactic
latitude component of the High Time Resolution Universe survey, as
recently done for the intermediate latitude
component~\citep{psj+14}. Such a search is in fact currently underway
(Champion et al. in prep.). We also estimate that \sarahfrb was twice
as bright as initially reported. Some of this difference is
attributable to the use of the full-width at half maximum by
\citet{bb14}, with a factor of $\sqrt{2}$ discrepency in the S/N which
we postulate to be due to the root 2 problem. 

In addition to the measured parameters we can determine a number of
model-dependent quantities, and several of these are listed in
Table~\ref{tab:FRB_vals_models}. Throughout we include \kb in our
analyses but note that this event has a very high probability of being
Galactic in origin~\citep{kskl12,bm14}: we use this to gauge a
reasonable uncertainty in the maximum Milky Way contribution to each
event's DM. Subtracting this contribution one can invoke a model
relating DM and redshift~\citep{ioka03,inoue04} to derive an upper
limit redshift estimate for each event. Unlike \citet{tsb+13} we do
not subtract a putative host contribution as it adds to the number of
assumptions needed (i.e. that there is a galactic host and that we
must assume a knowleadge of its composition). We also ignore the
variance in the IGM-contributed DM (see \citealt{mcq14}), as, without
independent redshift measurements, this has not yet been
quantified. Based on the therefore rather crude redshift estimate we
can derive comoving and luminosity distances in the standard
cosmology~\citep{wri06}, and thence the energy released to produce the
radio burst. The estimates for the energy cover a range of 2 orders of
magnitude. Determining the rest-frame peak luminosity for both the
upper limit on the intrinsic width (the observed width less the
effects of dispersion smearing and, where observed, multi-path
scattering) and a rest-frame timescale does not lessen this large
range of values. The unknown angular offsets (which, if accounted for
would boost the flux density and fluence), and the likely large
variance from the average DM as a function of redshift, neither of
which have we accounted for, will only increase the range. Thus
\textit{we do not have enough precision in our measurements to
  determine if the FRBs are standard candles or not}, even assuming
the models we have applied here are accurate. As all values quoted in
Table~\ref{tab:FRB_vals_models} are highly model-dependent and subject
to large uncertainties we suggest caution when using these values for
population analyses.

\begin{table*}
  \begin{center}
    \caption{\small{This table summarises some of the model-dependent
        parameters derived from the measured values presented in
        Table~\ref{tab:FRB_vals_measured}. DM$_{\mathrm{MW}}$ is the
        maximum Milky Way contribution to the DM for the line of sight
        to the FRB according to the NE2001 model of the Galaxy's free
        electron content~\citep{cl02}. The uncertainties on
        DM$_{\mathrm{MW}}$ are taken to be a factor of $2$, enough to
        imply a Galactic origin for \kb, as seems to be
        realistic~\citep{kskl12,bm14}. This uncertainty propagates
        through to all the following derived parameters. The ratio of
        DM to DM$_{\mathrm{MW}}$ is also given, where a value greater
        than $1$ nominally implies an extragalactic source. The
        redshift value is simply determined from $z_{\mathrm{model}} =
        (DM-DM_{\mathrm{MW}})/1200$~\citep{ioka03,inoue04}. The
        comoving distance is determined using the calculator of
        \citet{wri06}. The luminosity distance is simply $(1+z)$ times
        the comoving distance. The energy is calculated as the product
        of the observed fluence, the blueshifted effective bandwidth
        of the observations, and the square of the luminosity
        distance. No assumption of isotropy is made, i.e. a beaming
        solid angle of 1 steradian is used as per \citet{tsb+13}.}}
    \label{tab:FRB_vals_models}
    \setlength{\extrarowheight}{3 pt}
    \begin{tabular}{ccccccccc}

      \hline\hline Event & DM$_{\mathrm{MW}}$ & DM/DM$_{\mathrm{MW}}$ & $z_{\mathrm{model}}$ & $D_{\mathrm{comov}}$ & $D_{\mathrm{L}}$ & Energy \\
       & ($\rm{cm^{-3}\,pc}$) & &  & (Gpc) & (Gpc) & ($10^{32}$ joules) \\
      \hline
      Parkes FRBs & & & & & & &  \\
      \sarahfrb$^{\dagger}$  &  110 &  7.2  & $<0.57^{+0.04}_{-0.05}$   & $<2.1^{+0.12}_{-0.16}$     & $<3.4^{+0.3}_{-0.4}$  & $<2.5^{+2.1}_{-1.2}$ \\
      \kb$^{\dagger}$        &  523 &  1.4  & $<0.19^{+0.21}_{-0.19}$   & $<0.8(8)$                & $<0.9^{+1.3}_{-0.9}$  & $<0.1^{+1.3}_{-0.1}$ \\
      \lb$^{\dagger}$        &  45  &  8.4  & $<0.28^{+0.01}_{-0.02}$   & $<1.13^{+0.04}_{-0.08}$  & $<1.4(1)$             & $<2.1$ \\
      \tone$^{\clubsuit}$    &  35  &  27.2 & $<0.76^{+0.01}_{-0.02}$   & $<2.71^{+0.03}_{-0.06}$  & $<4.8^{+0.1}_{-0.2}$  & $<9.4^{+0.6}_{-4.2}$ \\
      \ttwo$^{\clubsuit}$    &  48  &  15.2 & $<0.56(2)$                & $<2.10^{+0.07}_{-0.06}$  & $<3.3^{+0.1}_{-0.2}$  & $<0.5^{+0.5}_{-0.3}$ \\
      \tthree$^{\clubsuit}$  &  32  &  34.1 & $<0.89^{+0.02}_{-0.01}$   & $<3.07^{+0.05}_{-0.03}$  & $<5.8(1)$             & $<3.6^{+3.2}_{-2.5}$ \\
      \tfour$^{\clubsuit}$   &  32  &  17.4 & $<0.43^{+0.02}_{-0.01}$   & $<1.67^{+0.07}_{-0.03}$  & $<2.4(1)$             & $<0.2(1)$ \\
      \emilyfrb$^{\clubsuit}$ & 35  &  16.1 & $<0.44(1)$                & $<1.71^{+0.03}_{-0.03}$  & $<2.46^{+0.04}_{-0.06}$ & $<0.4^{+0.4}_{-0.2}$ \\ \\
      Arecibo FRB & & & & & & & \\
      \spitlerfrb            &  188 &  3.0  & $<0.31(8)$                & $<1.2(3)$              & $<1.6^{+0.5}_{-0.4}$  & $<0.133^{+0.674}_{-0.127}$  \\

    \end{tabular}
  \end{center}
\end{table*}

\section{Selection Effects in FRB Searches}\label{sec:incompleteness}
%As well as the selection effects discussed above which hinder
%detection there are a number which should be considered when
%interpreting the detections and attempting to derive population
%statistics. 
The estimate for FRBs detectable by the current setup at Parkes is
$\sim 10^4$ FRBs/($4\pi$~sr)/day~\citep{tsb+13}. This number is simply
the observed rate of 4 FRBs in 23 days, extrapolated, from the $\sim
0.55\deg^2$ half-power field-of-view of the multi-beam receiver at
Parkes~\citep{swb+96}, to the entire sky. In addition to the obvious
caveats of such an extrapolation, the meaning of this rate must be
interpreted carefully, for a number of reasons.

\begin{figure}
  \begin{center}
    % l b r t
    \includegraphics[scale=0.35,trim = 15mm 20mm 15mm 15mm, clip, angle=0]{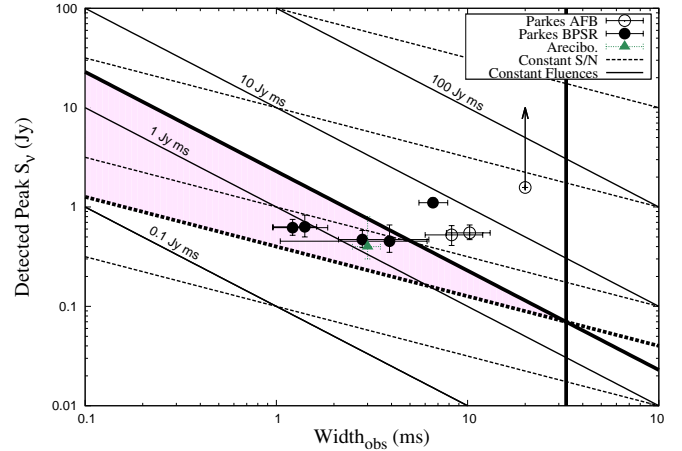}
    \caption{\small{The detected flux density-observed width parameter
        space. Lines of constant signal-to-noise (dashed) and constant
        fluence (solid) are shown. Events above the thick dashed line
        are detectable at Parkes. In this case, and if all FRBs are
        less than some maximum putative width (here denoted by the
        thick vertical line), we only have fluence completeness above
        $\sim2\;\mathrm{Jy}\,\mathrm{ms}$. In the shaded triangle we
        have fluence incompleteness. The Arecibo FRB (green triangle)
        is included for illustration only: different incompleteness
        regions apply to different observing configurations.}}
%Here we have used the High Time Resolution Universe survey limits
%which are slightly deeper than the archival survey as the bandwidth
%is slighter wider and the digitisation is 2 bit.
 \label{fig:completeness}
  \end{center}
\end{figure}

\textit{Fluence \& Width Incompleteness?}: One might 
%na\"{i}vely
suggest that the all-sky rate is that above the flux-density threshold
at the half-power beam-width of Parkes. 
%Using the system configuration
%relevant to the data wherein the 3 bursts discussed here were found
%this corresponds to flux densities of $1.01/\sqrt{W/{\rm ms}}$~Jy
%($1.07/\sqrt{W/{\rm ms}}$~Jy or $1.27/\sqrt{W/{\rm ms}}$~Jy) for the
%central (inner or outer ring) receivers, for an $8$-$\sigma$ signal.
% This calculation uses a BW of 288 MHz (the PKS analog FB BW), Tsys
% of 28K (Parkes Observing Guide), 1-bit digitisation and gains of
% 0.735, 0.690 and 0.581 K/Jy for the central, inner ring and outer
% ring beams respectively (mlc+01).
%
%   Smin = (8)*(Tsys/G)*beta*(np*BW*width)^-0.5
%   beta = 1/0.798
% -> Smin = 370/G mJy (W/ms)^-0.5
%
% Smin = 503 mJy/ sqrt(W/ms)   CENTRAL BEAM
% Smin = 536 mJy/ sqrt(W/ms)   INNER RING
% Smin = 636 mJy/ sqrt(W/ms)   OUTER RING
% and twice these values at the half-power point
%
Even though the physically relevant parameter is the fluence, i.e. the
area under the pulse curve in the dedispersed time series, our
sensitivity depends also on the pulse width. Thus, pulses with the
same fluence but different widths (due to traversing different paths
in the IGM/ISM) are not equally
detectable. Figure~\ref{fig:completeness} shows the sensitivity to
bursts in the flux density-width plane. We are always incomplete to
wide bursts. However if FRBs do not exist above some maximum
width\footnote{The width of $32$~ms used in
  Figure~\ref{fig:completeness} is the maximum to which FRB searches
  at Parkes are sensitive~\citep{psj+14}.} we are still left with an
incomplete sampling of fluence. This feeds into the often posed
question of `what is the $\log N-\log S$ distribution?'. In this case
`S' ought to be fluence, and fluence completeness should be accounted
for. Considering Figure~\ref{fig:completeness} this would imply
discarding half of the Parkes FRBs and then binning the
remainder. With such a small number of detections this is not yet a
meaningful exercise. Fluence completeness should be considered when
determining population estimates.
%, especially when
%extrapolating to lower frequencies where observed pulses are likely to
%be wider~\citep{ttw13,clh+14}.

\textit{Latitude dependence?}: Some additional selection effects have
been identified empirically, e.g. FRBs seem to be much more difficult
to detect at lower Galactic latitudes, despite intensive
efforts~\citep{psj+14,bb14}. This points towards a Galactic
obscuration effect but at present none of the known possibilities are
sufficient to explain the paucity of low-latitude detections. The
possibility remains that the rate (extrapolated from 4 events, see
above) is in fact too high.

\textit{What is an FRB?}: There is also the question as to how to
algorithmically define what an FRB is, in contrast to a ``RRAT'':
pulsars which show detectable single pulses of radio emission only
very occasionally~\citep{km11}. If we detect a single pulse from a
RRAT how do we distinguish it from an FRB? A RRAT may eventually
repeat but the main difference is that RRATs are Galactic whereas FRBs
are believed to be extragalactic. We decide upon this based on the
ratio of the $DM$ of a detected event to the maximum Galactic $DM$
contribution along the line of sight, $DM_{\mathrm{MW}}$. If
$x=DM/DM_{\mathrm{MW}}>1$ then the source is extragalactic, but the
model we use to determine $DM_{\mathrm{MW}}$ is quite uncertain,
especially at high Galactic latitudes~\citep{cl02}. This has resulted
in \textit{ad hoc} selection rules such as selecting events with
$x>0.9$~\citep{psj+14}, but, as noted in \citet{sch+14}, there is a
rather wide `grey area' in this parameter space. As the uncertainty on
$x$ depends both on the $DM$ and the line of sight in very asystematic
ways this is difficult to quantify and it is quite conceivable that
many FRBs have already been detected and falsely labelled as RRATs,
and the converse may already be the case for \kb. In effect there is a
low-redshift blindness to FRBs, analogous to the low-$DM$ blindness in
Galactic SP searches~\citep{kle+10}.

\section{Conclusions}\label{sec:last}
With so many selection effects evident, large uncertainties in flux
density and fluence estimates (see \S~\ref{sec:archival_FRBs}),
essentially no knowledge of FRB spectra, and when dealing with such
small number statistics, serious population analyses are
precluded. These obstacles will be overcome only when a much larger
number of FRBs are discovered. For example, to remove the fluence
incompleteness one could simply discard all FRBs below the
incompleteness value. Considering the Thornton events, and under the
assumption that no FRBs are broadened more than $\approx 32$~ms, this
means discarding three out of the four events and interpreting the
Parkes FRB rate above a fluence of $\sim 2\;\mathrm{Jy}\,\mathrm{ms}$
to be $\sim 2500$ per sky per day. Clearly this estimate is hugely
uncertain and such analyses will only become more practical (see
Figure~\ref{fig:completeness}) when a much larger sample is obtained
and the population below the incompleteness boundary can be
modelled. To remove the low-redshift blindness we might benefit from
independent distance estimates. New means of estimating the distance,
such as the method of \citet{bm14}, could shed some light on this if
they are applied to a large sample of RRATs and FRBs. We encourage the
community to search all ongoing and archival surveys for FRBs: given
the issues raised in \S~\ref{sec:SP_searches} we propose that this is
best done using \H, or one using the same search algorithm.
%The clear
%scientific potential of FRBs further motivates a search for more, and
%as such we are currently performing a large-scale search of the
%existing archival data, some of which has either never been searched
%previously, or has only been searched to a limited extent. We will
%present the results of this search in subsequent papers. 
Such searches could yield a few tens of new FRBs. Beyond that the only
way to discover hundreds to thousands of FRBs is to use high
sensitivity wide field-of-view telescopes with a large amount of
on-sky time, e.g. through `piggy-back' transient observations with
MeerKAT and SKA1-Mid~\citep{rob15}. This will be necessary to maximise
the scientific return on yet-to-be detected FRBs~\citep{pbb+14}.

%Furthermore we can use the multiple detections of \lb to calculate a
%more refined sky position for the source, by noting the beam
%configuration and invoking a simple Gaussian beam
%model~\citep{swb+96}. It is straightforward to show that for two beams
%with relative \textit{detected} flux densities $r$, full-width
%half-maxima $f$ and beam separation $s$, the angular offset to the
%true position of the source, along the line joining the two beams, is
%given by: $s/2 + (1/s)(f/2.355)^2\ln r$. Applying this method to the
%\lb detections defines a region on the sky whose centre is at
%$(\alpha, \delta)=(19.154, -75.155)$, where $\alpha$ is in degrees of
%right ascension and $\delta$ is in degrees of arc. The offset with
%respect to the centre of beam 6 is comparable to the error in this
%position so that it seems clear that \lb occurred within the the
%full-width half-power radius of beam 6 towards the WSW direction from
%the centre of the beam.
%%Considering that only a lower limit on the S/N value in
%%beam 6 can be made, as the digitisers were saturated by the event,
%%moves the derived position closer to the centre of beam 6. 
%As no other FRBs were detected in more than one beam no such
%positional refinement is possible for any of the other events. Their
%non-detection in surrounding beams is also quite poorly constraining,
%typically limiting the burst no better than within $\sim10$~arcmin of
%the centre of the beam in which they were detected.

\section*{Data \& Software Release}
We have made the data easily available for the FRBs discussed in this
paper. The data can be accessed via the Research Data Australia
Portal\footnote{\texttt{http://researchdata.ands.org.au/fast-radio-bursts-parkes/}}.
%The data can be accessed via a \textsc{dropbox}
%repository\footnote{\texttt{http://tinyurl.com/frb-data}} and will
%published more formally in the coming months.
%\footnote{\texttt{https://www.dropbox.com/sh/l6kr87oxsxpg367/AAAqdwv-ZdqQcMpXoZmfK0G3a}}. Additionally
Additionally analysis software is available and can be accessed via a
\textsc{github}
repository\footnote{\texttt{https://github.com/evanocathain/Useful\_FRB\_stuff}}. These
have been used in the preparation of this paper. It is our hope that
others can use these to directly access and analyse the raw data
collected at the telescope. In this way uncertainties and
misinterpretations can be minimised, and new predictions and analysis
tools can be quickly tested.

\section*{Acknowledgements}\label{sec:public_data}
%This work used the gSTAR national facility which is funded by
%Swinburne and the Australian Government’s Education Investment
%Fund. 
%The Parkes Radio Telescope is part of the Australia Telescope
%National Facility, which is funded by the Commonwealth of Australia
%for operation as a National Facility managed by CSIRO. 
We thank E. Barr, J-P. Macquart and an anonymous referee for extensive
helpful comments which have hugely improved the quality of this
manuscript.
%, and acknowledges the FSM while
%cursing Xenu. 
EFK and EP acknowledge the support of the Australian Research Council
Centre of Excellence for All-sky Astrophysics (CAASTRO), through
project number CE110001020.

%%                  %%
%% THE BIBLIOGRAPHY %%
%%                  %%
%\newpage
%\input{paper.bbl}
%\bibliography{journals,journals_apj,psrrefs,modrefs,crossrefs}
\bibliographystyle{mnras}

\end{document}